\documentclass{kapprocmod}
\usepackage{graphicx}
\let\footnote\savefootnote
\let\footnotetext\savefootnotetext
\let\footnoterule\savefootnoterule
\kluwerbib
\startauthorindex
\newcommand{\EQ}[1]{\begin{equation}\label{#1}}
\newcommand{\EN}{\end{equation}}
\newcommand{\Fig}[1]{Fig.\ \ref{#1}}

\def\Rm{R_{\rm m}}
\def\b{{\rm b}}
\def\apj{Astrophysical Journal}

\def\solphys{Solar Physics}
\begin{document}
\graphicspath{{Nordlund/}}
%\thispagestyle{empty}
%
%%%%%%%%%%%%%%%%%%%%%%%%%%%%%%%%%%%%%%%%%%%%%%%%%%%%%%%%%%%
%%%%%%%%%%%%%%% ARTICLE TITLE %%%%%%%%%%%%%%%%%%%%%%%%%%%%%
%%%%%%%%%%%%%%%%%%%%%%%%%%%%%%%%%%%%%%%%%%%%%%%%%%%%%%%%%%%
%%
%% Note the use of lower and upper case. In [] is the title for the
%% list of contents. In {} is the title at the top of your paper.
%% You can use \\ within {} but not within []
%% Edit the lines below
%%
\articletitle[Magnetic dissipation: Spatial and temporal structure]
{Magnetic dissipation:\\ Spatial and temporal structure}
%%
%%%%%%%%%%%%%%%%%%%%%%%%%%%%%%%%%%%%%%%%%%%%%%%%%%%%%%%%%%
%%%%%%%%%%%%%%  RUNNING HEADS %%%%%%%%%%%%%%%%%%%%%%%%%%%%
%%%%%%%%%%%%%%%%%%%%%%%%%%%%%%%%%%%%%%%%%%%%%%%%%%%%%%%%%%
%%
%% Supply the author(s) and the title for the running head:
%% Check if they fit in one line when printed out.
%%
\booktitlerunninghead{{\AA}.~Nordlund}
\chaptitlerunninghead{Magnetic dissipation: Spatial and temporal
structure}
%%
%%%%%%%%%%%%%%%%%%%%%%%%%%%%%%%%%%%%%%%%%%%%%%%%%%%%%%%%%%
%%%%%%%%%%%%%%  AUTHORS   %%%%%%%%%%%%%%%%%%%%%%%%%%%%%%%%
%%%%%%%%%%%%%%%%%%%%%%%%%%%%%%%%%%%%%%%%%%%%%%%%%%%%%%%%%%
%%
%% Insert names of the authors below (capital letters).
%% \thanks makes a footnote
%% Mathstyle superscripts, e.g., $^1$, are used for affiliation
%% Supply e-mail address of the corresponding author.
%% If there is a footnote to the corresponding author's address you
%% can put his e-mail address there, otherwise put \email{<address>}
%% immediately after \affil{} of the corresponding author.
%% Example:
%%
%% %%%%%%%%%%%  First author   %%%%%%%%%%%%%%%%%%%%%%%%%%%%%%
%%
\author{{\AA}KE NORDLUND\nobreak
%%
%% If the address for correspondence of the corresponding author is
%% different from his current affiliation then make a footnote (which
%% includes his e-mail address). The command for this footnote is
%% \thanks :
%%
%\thanks{Address for correspondence: Warsaw
%University, Institute of Geophysics, ul. Pasteura 7, 02-093 Warszawa,
%Poland; \tt{kbajer@fuw.edu.pl}}\nobreak
%%
%%
%%%%%%%%%%%%%%%%%%%%%%%%%%%%%%%%%%%%%%%%%%%%%%%%%%%%%%%%%%%%%%%
}\vskip6pt
%
%% %%%%%%   FIRST AFFILIATION  %%%%%%%%%%%%%
%
\affil{
%\hspace*{15pt}
Niels Bohr Institute and Theoretical Astrophysics Center,\\
%\hspace*{20pt}                  % DELETE IF SINGLE AFFILIATION
Juliane Maries Vej 30, DK-2100 Copenhagen {\O}, Denmark
%%
%%%%%%  DELETE THE FOLLOWING 2 LINES IF YOU GAVE THE
%%%%%%  CORRESPONDING AUTHOR'S E-MAIL ADDRESS IN A FOOTNOTE
\\ \hspace*{22pt}
{\emailfont aake@astro.ku.dk}
}\vskip6pt
%%
%%%%%%%%%%%%%%  ADDRESS LABELS   %%%%%%%%%%%%%%%%%%%%%%%%%
%%
%% Please give here accurate addresses for correspondence of ALL
%% authors following the example below. Leave the addresses commented
%% out, they will not appear in your paper but will appear in the
%% `List of Participants' to be included in the Proceedings.
%%
%% Dr. {\AA}. Nordlund              \e
%% Astronomical Observatory         \e
%% Juliane Maries Vej 30            \e
%% DK-2100 Copenhagen {\O}          \e
%% DENMARK                          \e
%% \tel{+45 3532 5968}              \e
%% \fax{+45 3532 5989}              \e
%% \email{aake@astro.ku.dk}         \e
%% \url{www.astro.ku.dk/~aake}
%%
%%%%%%%%%%%%%%%%%%%%%%%%%%%%%%%%%%%%%%%%%%%%%%%%%%%%%%%%%%

%% Just before the abstract of your paper write other relevant keywords:
%% \inxx{topology of the magnetic field}
%% \inxx{linkage of the magnetic flux tubes}

\inxx{magnetic dissipation}
\inxx{coronal heating}
\inxx{fast reconnection}
\inxx{numerical experiments}
\inxx{3D mhd models}
\inxx{magnetic field line braiding}
\inxx{synthetic emission measures}

\vspace*{-8pt}\begin{abstract}

A magnetically dominated plasma driven by motions on boundaries
at which magnetic field lines are anchored is forced to
dissipate the work being done upon it, no matter how small the
electrical resistivity. Numerical experiments have clarified the
mechanisms through which balance between the boundary work and
the dissipation in the interior is obtained. Dissipation is
achieved through the formation of a hierarchy of electrical
current sheets, which appear as a result of the topological
interlocking of individual strands of magnetic field. The
probability distribution function of the local winding of
magnetic field lines is nearly Gaussian, with a width of the
order unity.  The dissipation is highly irregular in space as
well as in time, but the average level of dissipation is well
described by a scaling law that is independent of the electrical
resistivity.

If the boundary driving is suspended for a period of time the
magnetic dissipation rapidly drops to insignificant levels,
leaving the magnetic field in a nearly force-free, yet
spatially complex state, with significant amounts of free
magnetic energy but no dissipating current sheets. Renewed
boundary driving leads to a quick return to dissipation levels
compatible with the rate of boundary work, with dissipation
starting much more rapidly than when starting from idealized
initial conditions with a uniform magnetic field.

Application of these concepts to modeling of the solar corona
leads to scaling predictions in agreement with scaling laws
obtained empirically; the dissipation scales with the inverse
square of the loop length, and is proportional to the surface
magnetic flux.  The ultimate source of the coronal heating is
the photospheric velocity field, which causes braiding and
reconnection of magnetic field lines in the corona. Realistic,
three-dimensional numerical models predict emission measures,
coronal structures, and heating rates compatible with
observations.

\end{abstract}
\vspace*{-6pt}
\noindent\rule{\textwidth}{0.01mm}
\vskip12pt\hfill\hspace*{-12pt}\begin{tabular}[t]{r}
{\it\small\begin{tabular}{l}
When driv'n by extreme agitation,\\
I am subject to fierce dissipation;\\
\quad In each current sheet\\
\quad There's created much heat,\\
And my field thus achieves saturation.
\end{tabular}}
\end{tabular}
\nobreak\hspace*{12pt}

%%%%%%%%%%%%%%%%%%%%%%%%%%%%%%%%%%%%%%%%%%%%%%%%%%%%%%%%%%%%%%%%
%%%%%%%%%%%%    BODY OF ARTICLE    %%%%%%%%%%%%%%%%%%%%%%%%%%%%%
%%%%%%%%%%%%%%%%%%%%%%%%%%%%%%%%%%%%%%%%%%%%%%%%%%%%%%%%%%%%%%%%

%% We are very keen to produce a useful index to this volume.
%% As you write the text of your paper, please mark every specific term
%% you use with \inx{} macro. For example you write a sentence:
%% "\inx{Magnetic helicity}, for instance, does not
%% distinguish the \inx{Borromean rings} from unlinked rings."

\newpage
\section{Introduction}

Magnetic fields are ubiquitous in astrophysical objects;
circumstances where there is no magnetic field present are
exceptions. Indeed, much of the non-thermal activity that is
observed in astrophysical systems is probably related to the
presence of magnetic fields.

Gravity is, indirectly, a major reason for the ubiquitous
magnetic activity, because it tends to separate matter into
dense and tenuous regions.  Magnetic fields that connect such
regions are subjected to stress in the dense regions, and are
forced to dissipate in the tenuous regions.  There, the magnetic
field energy density can be many times higher than the thermal
and kinetic energy density of the gas, and minor readjustments
of the magnetic field may correspond to significant heating and
acceleration of the gas.

Understanding the principles that control the dissipation of
magnetic energy when the \inx{plasma beta} ($\beta=P_g/P_B$,
where $P_g$ is gas pressure and $P_B$ is magnetic pressure)
is low and the magnetic \inx{Reynolds number} ($\Rm = U L /\eta$,
where $U$ is velocity, $L$ size, and $\eta$ magnetic diffusivity)
is very high is a major challenge, and numerous research papers,
review articles and books have been published on this subject
over the years
(e.g., Parker 1972, 1983, 1988, 1994;
\cite{Sturrock+Uchida81};
\cite{vanBall86};
\cite{Mikic+ea89};
\cite{Heyvaerts+Priest92};
\cite{Longcope+Sudan94};
\cite{Galsgaard+Nordlund96a}b;
\cite{Nordlund+Galsgaard97};
\cite{Gomez+00};
to mention just a few).

The solar corona is an ideal `test site' for theories and models
of magnetic dissipation, since it provides rich opportunities
for observing both the spatial and temporal structure of a
dissipating low beta plasma. The Sun is indeed a
`\inx{Rosetta stone}' in the context of magnetic dissipation---once we
understand how magnetic dissipation occurs under such well
observed conditions we may be much more confident when
extrapolating to more distant and less well observed
circumstances.

Numerical experiments have emerged as a complementary and rich
source of inspiration in the quest to understand magnetic
dissipation.
%dissipation---several of the references cited above report
%results from such experiments.
Below I briefly summarize
conclusions from two types of experiments; generic experiments
where a low beta plasma is driven from two opposing boundaries,
and realistic experiments that attempt to model solar coronal
conditions as closely as possible.

\section{Boundary driven magnetic dissipation}

%% Fig. 1
\begin{figure}[t]
\begin{center}
 \includegraphics[height=5.1cm,width=5.2cm]{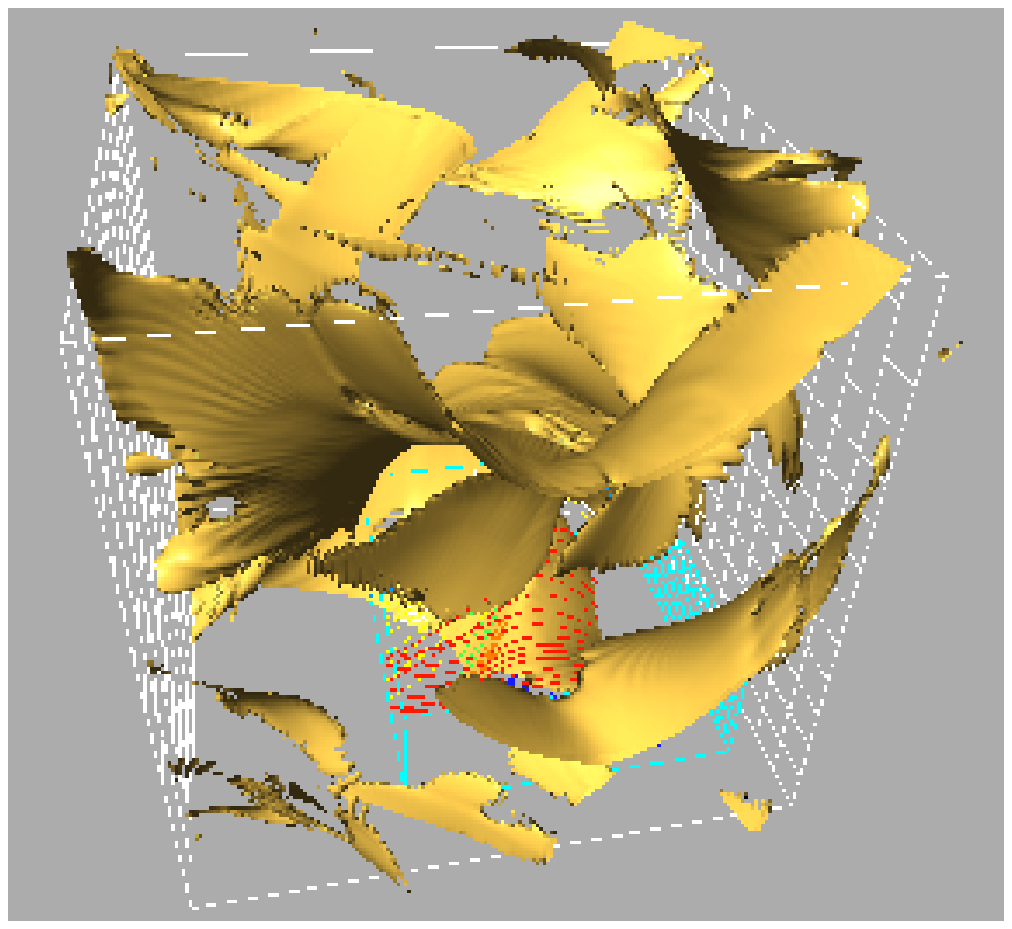}
 \hspace*{-0mm}({\it a})
 \hspace*{0mm}
 \includegraphics[height=5.1cm,width=5.2cm]{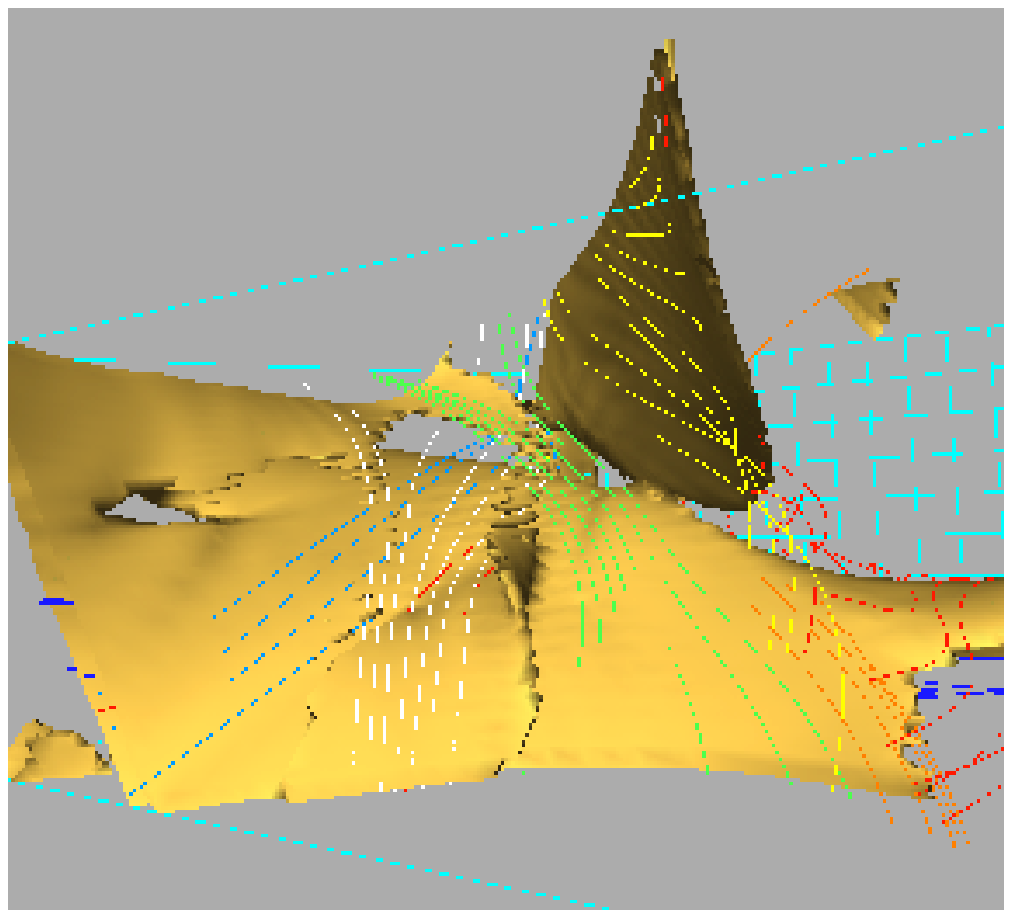}
 \hspace*{-0mm}({\it b})
\end{center}
\label{fig1-nordlund}
\caption[]{a) Hierarchy of current sheets, shown as
\protect{\inx{isosurfaces}} of constant electric current density
(\cite{Galsgaard+Nordlund96a}).  b) Detail of the hierarchy.}
\end{figure}

Generic experiments demonstrate that \inx{magnetic dissipation}
does not occur in simple, \inx{monolithic current sheets}, even
if the \inx{boundary motions} are large scale and slow
(cf.\ \cite{Galsgaard+Nordlund96a} for details about such
experiments).  Rather, a \inx{hierarchy of
current sheets} form, with smaller scale current sheets
protruding from larger scale ones (cf.\ Fig.\ 1).

\begin{figure}[t]
%\hspace*{-3mm}
\includegraphics[height=5.3cm,width=6.0cm]{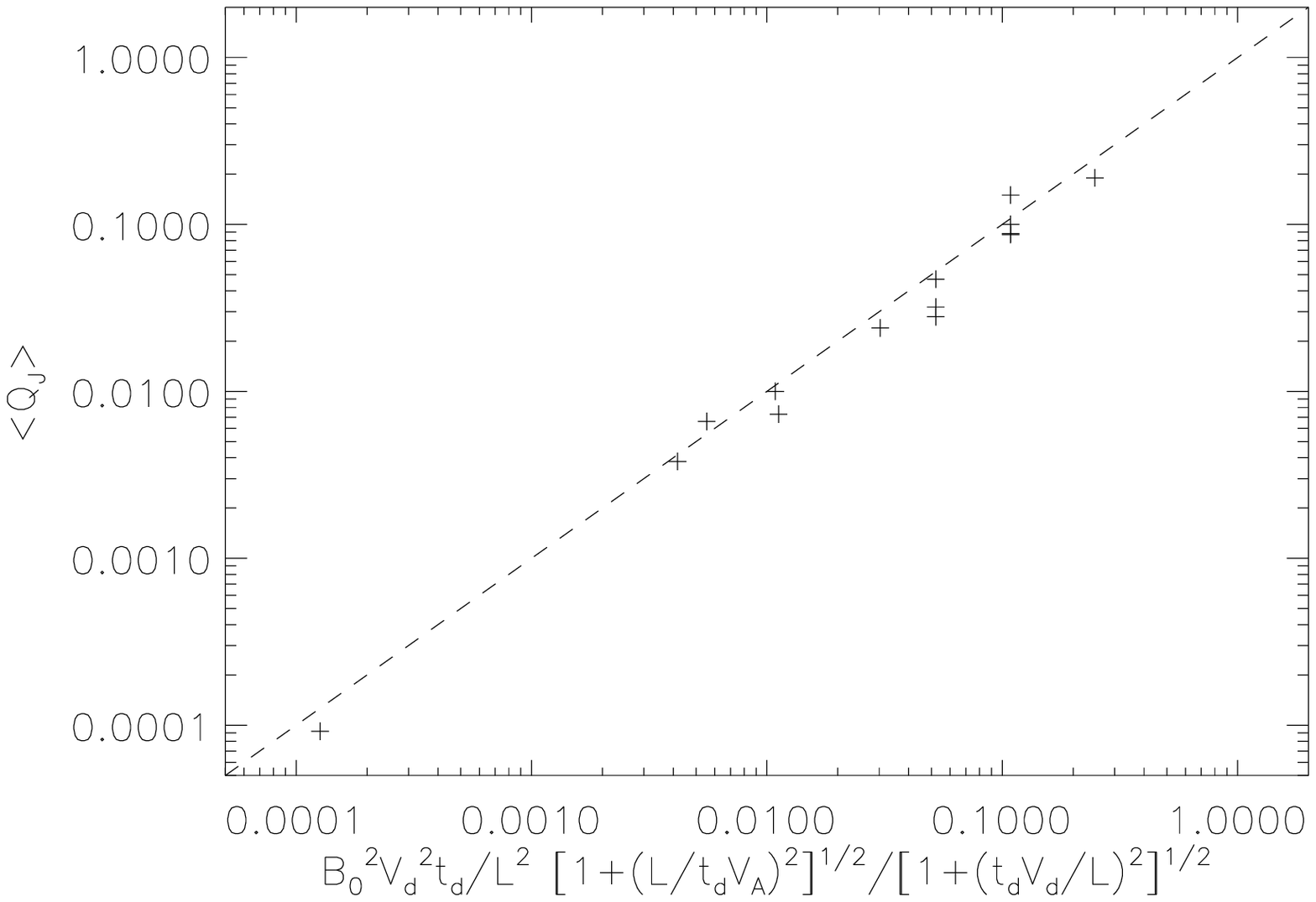}
\hspace*{-6mm}\vspace*{-12mm}({\it a})
\hspace*{-4mm}\vspace*{+12mm}
\includegraphics[height=5.3cm,width=6.2cm]{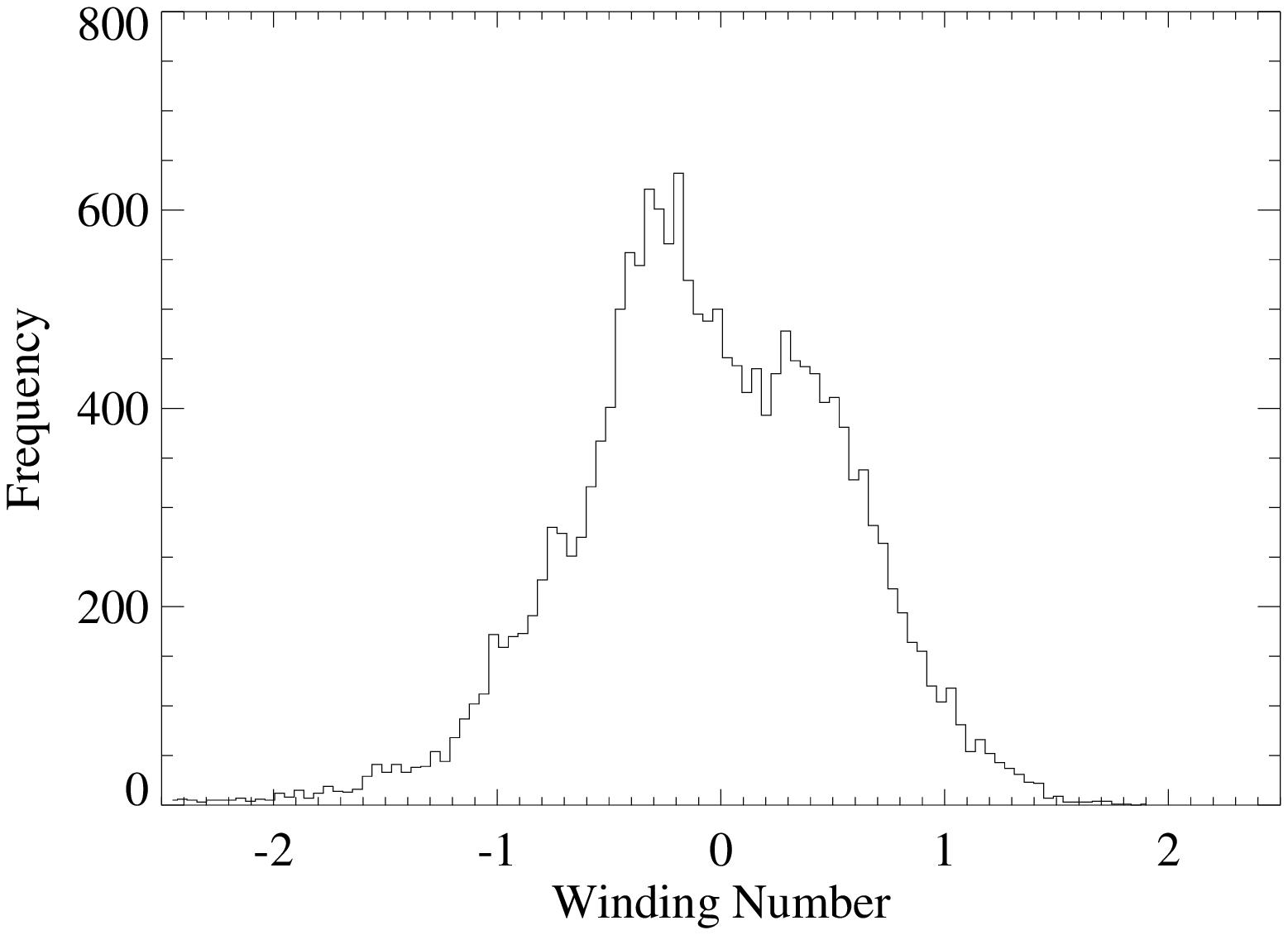}
\hspace*{-6mm}\vspace*{-12mm}({\it b})
\hspace*{+0mm}\vspace*{+12mm}
\vspace*{-5mm}
\caption[]{
a) Scaling law for magnetic dissipation
(\cite{Galsgaard+Nordlund96a}).
b) Distribution of winding numbers
(\cite{Nordlund+Galsgaard97}).
}
\label{fig2-nordlund}
\end{figure}

On the average, the \inx{collective dissipation} in the hierarchy of
current sheets balances the boundary work, as it must, to
satisfy \inx{energy conservation}. The average level of work and
dissipation at which the equilibrium is obtained does not depend
noticeably on the \inx{resistivity} (or, equivalently, the \inx{numerical
resolution}), as long as the \inx{magnetic Reynolds number} is not too
small. The work $W$, and hence the dissipation $Q$ times the distance
$L$ between the boundaries, is proportional to the energy
density of the magnetic field at the boundary, the \inx{average angle
of inclination} $\phi$ of the magnetic field at the boundary, and
the \inx{boundary velocity} $v_\b$.
\EQ{scaling}
W_{\b} = Q L \sim {B^2 \over 8\pi} \ \tan\phi \ v_{\b}
\EN
The crucial angle factor $\tan\phi$ scales as $\tan\phi \sim
{v_{\b} \tau_{\b} / L} \sim {\ell_{\b} / L}$, where
$\tau_{\b}$ is the \inx{autocorrelation time} of the boundary motions,
$\ell_{\b}$ is the ``\inx{stroke length}'' of the boundary motions,
and $L$ is the distance between the driving boundaries. The
\inx{average dissipation} thus scales as
\EQ{scaling}
Q \sim {B^2 \over 8\pi}\ {v_\b \ell_\b  \over L^2}
\label{nordlund-eq-scaling}
\EN
(\Fig{fig2-nordlund}a).  Incidentally, it turns out that the
\inx{scaling of the photospheric motions} that drive the solar
corona is such that $v_\b(\ell_\b) \sim 1/\ell_\b$, which
implies that similar contributions to the driving are obtained
from an extended range of scales.

%% Fig. 2

A sequence of uncorrelated strokes at the two opposing
boundaries would, if the \inx{connectivity} between the boundaries
was conserved, lead to further increase of the \inx{tilt angles}.
The magnetic field lines would become increasingly tangled,
as each new stroke would cause the end points of field lines
connected to neighboring points at one boundary to move
further apart at the other boundary.

But the \inx{development of current sheets} prevents connectivity from
being conserved, and prevents the build-up of angles much larger
than $\ell_{\b}/L$.  In the \inx{hierarchy of current sheets} the
smaller current sheets provide a \inx{dissipation path} for the larger
ones, down to the smallest current sheets at a few times the
\inx{resistive scale}. An increase of resolution (magnetic Reynolds
number) allows even smaller current sheets to form, but does not
change the large scale angles noticeably, and hence does not
influence the average level of dissipation.

In terms of the \inx{winding} of one field line around another, the
statistical steady state is one where the \inx{number of windings}
from one boundary to the other has an approximately \inx{Gaussian
distribution}, with a Gaussian width of order unity
(\Fig{fig2-nordlund}b).  It is indeed a well known
result that a twist much larger than unity leads to violent
instability (see \cite{Galsgaard+Nordlund96b} and references
cited therein).

%% Fig. 3
\begin{figure}[t]
%%AA Do NOT use centering on this figure
\includegraphics[height=8.5cm,width=12.5cm]{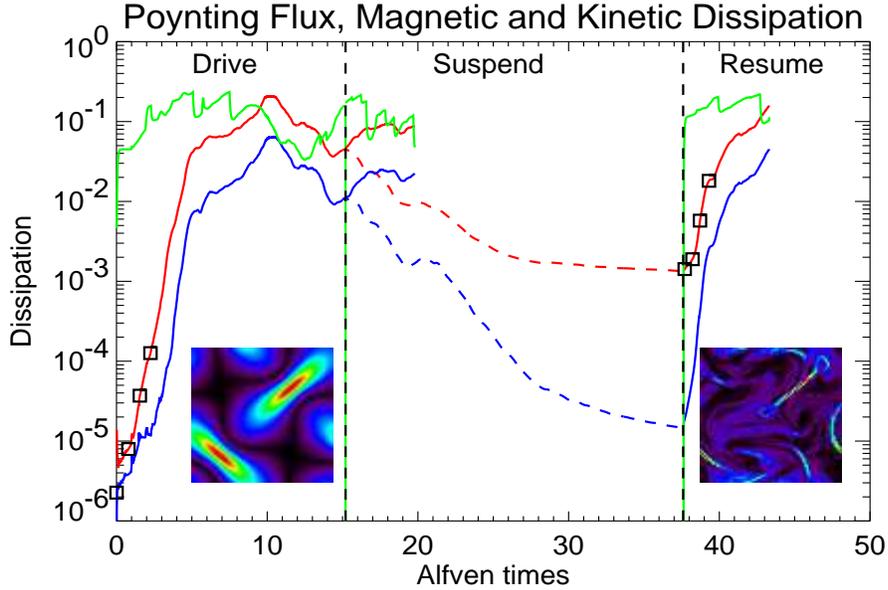}
\vspace*{-8mm}
\caption[]{The \inx{boundary work} (green/top), the \inx{magnetic dissipation}
(red/middle), and the \inx{kinetic energy dissipation} (blue/lower) in a \inx{3-D
numerical experiment} (Nordlund and Galsgaard, in preparation)
where a \inx{magnetically dominated plasma} is
driven by boundary work at two \inx{opposing boundaries}. At $t\sim
15$ the boundary work is turned off and the system is allowed to
relax until $t\sim 38$, where the initial \inx{velocity pattern} is
repeated again. The two insets show the \inx{electric current density}
in a cross section half-way between the two driving boundaries,
approximately two Alfv{\'e}n crossing times (defined as the length
of the box $L$ divided by the Alfv{\'e}n speed $B/\sqrt{4\pi\rho}$)
after the driving has started and resumed, respectively.}
\label{nordlund-fig3}
\end{figure}

\section{Suspended / resumed boundary driving}
\inxx{suspended driving} \inxx{resumed driving}

If the external work is suspended dissipation quickly drops, as
the current sheets die out. Distributed (smooth) currents
remain, but these dissipate much more slowly. The system is in
an approximately force-free state, similar to the one just
before current sheets first turned on. When the external work is
resumed current sheets return promptly, and dissipation quickly
returns to the previous average level (\Fig{nordlund-fig3}).

Note that, even though exactly the same velocity field is
applied for exactly the same time, there are much sharper
current concentrations (\inx{current sheets}) in the right-most inset
of \Fig{nordlund-fig3}, illustrating that the quiescent but
\inx{spatially complex magnetic field} in the \inx{suspended state} is
"ripe" for quickly producing current sheets.  The system thus
reaches balance between \inx{driving and dissipation} much more
quickly from the suspended driving ($\sim$ \inx{force free}) state
than from the initial (\inx{potential}) state.

%% Fig. 4
\begin{figure}[t]
\begin{center}
\includegraphics[height=5.2cm,width=5.2cm]
{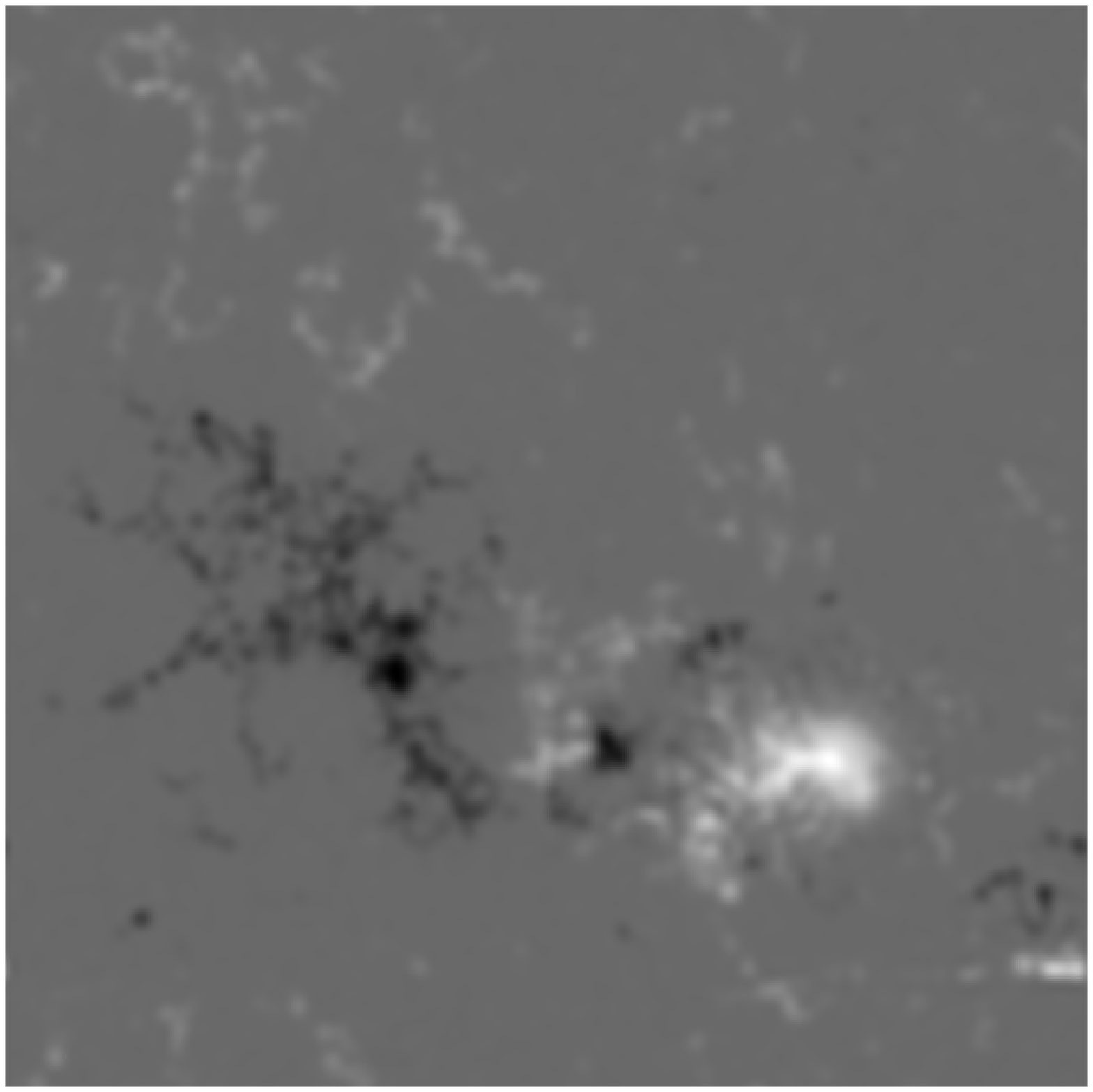}
\hspace*{-1mm}({\it a})
\hspace*{0mm}
\includegraphics[height=5.2cm,width=5.2cm]
{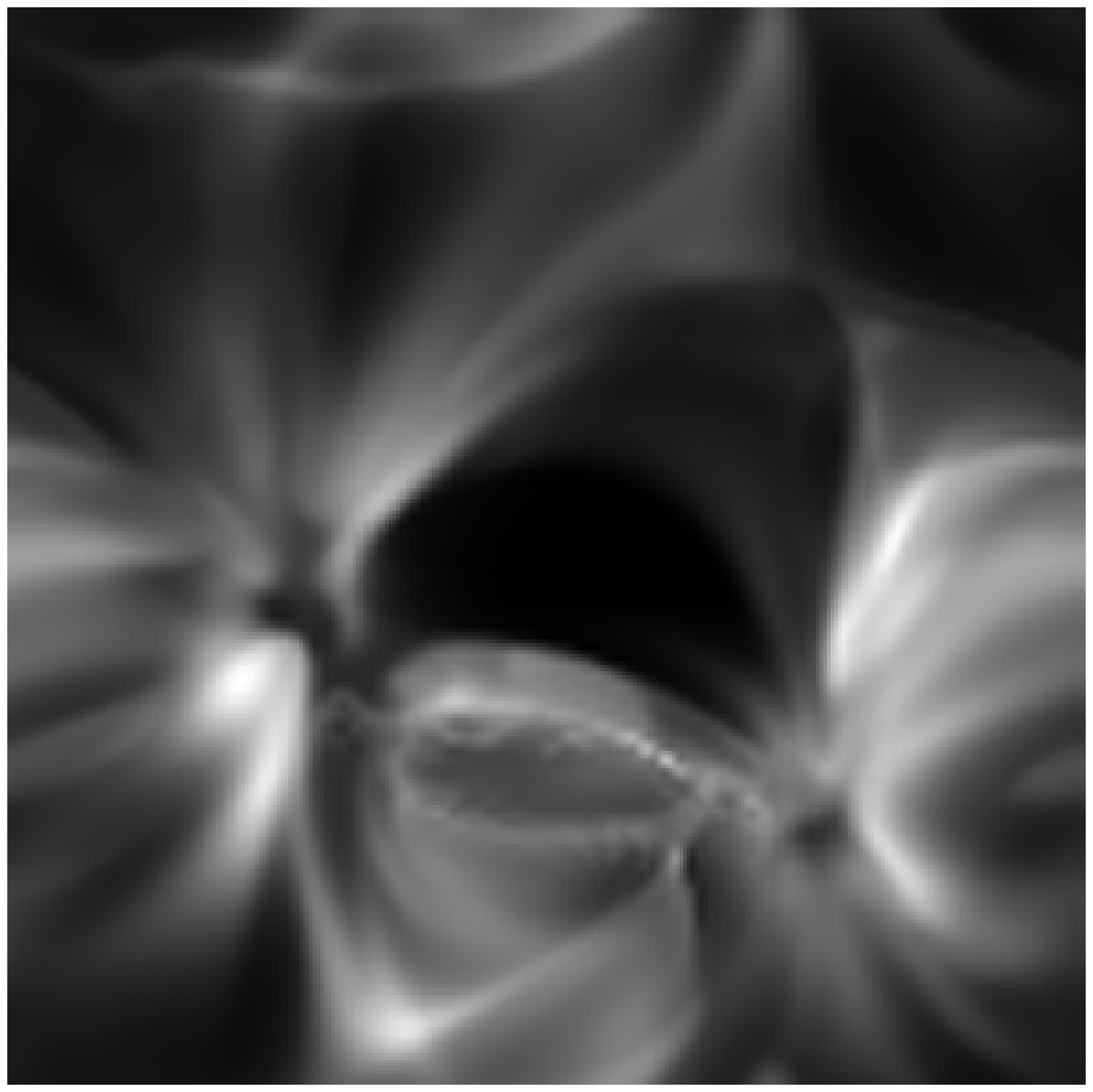}
\hspace*{-1mm}({\it b})
\end{center}
\caption[]{a) The initial condition for the photospheric magnetic
field; a high resolution magnetogram of active region 9114, from the
Michelson Doppler Interferometer instrument on board the SOHO
satellite.
b) Synthetic $\lambda$195 filter image.}
\vspace*{-24pt}
\label{nordlund-fig4}
\end{figure}

\section{Coronal heating experiments}

Realistic numerical models of \inx{coronal heating} and activity are
\inxx{coronal activity}
now within reach (\cite{Gudiksen+Nordlund02}), even though the
\inx{resolution constraints} are quite severe.  In order to apply
realistic and well calibrated \inx{boundary conditions} it is
necessary to bridge the distance from the base of the corona to
the \inx{photospheric surface}, where both the magnetic field and the
velocity field are known sufficiently well to specify the
\inx{boundary driving}. This limits the \inx{vertical resolution} to
better than, or of the order of, photospheric and chromospheric
pressure and \inx{density scale heights}.  The horizontal resolution
\inxx{pressure scale heights}
needs to be similar, to resolve the \inx{granular scales} where the
\inx{horizontal velocity} amplitudes are largest.   The size of a
model, on the other hand, needs to be of the order of the size of
an active region, or larger.
The very high \inx{Alfv{\'e}n velocity} in the corona above \inx{active
regions} limits (via the Courant condition) the \inx{time step}
to of the order of $10$--$30$ ms, at
the given \inx{spatial resolution}.  Time intervals of at least
several \inx{granulation turn over times} ($\sim 5$ min) need to be covered.
The resulting overall \inx{constraints} are demanding,
but not excessively so, given the power of current, \inx{massively
parallel supercomputers}.

With a relevant \inx{cooling function} approximation (Kahn 1976) and
\inx{Spitzer conductivity}, and by computing \inx{synthetic emission
measures} that correspond closely to those observed from, e.g.,
the \inx{TRACE satellite} (Aschwanden, Schrijver \& Alexander 2001),
one may compare the
results directly with solar conditions.  With \inx{initial
conditions} taken from a high resolution \inx{MDI magnetogram}
(\Fig{nordlund-fig4}a), and a random velocity field with scaling
properties consistent with a 5 km/s rms solar \inx{photospheric velocity
field}, one obtains \inx{average heating rates} over an active region
$\sim 10^7\ {\rm erg\,cm^{-2}\,s^{-1}}$, and \inx{emission measure
images} that are similar to observed ones (cf.\
\Fig{nordlund-fig4}b).

\section{Discussion and concluding remarks}

The results of the numerical experiments, and the properties of
the scaling law derived from them, provide evidence that we are
finally approaching a basic understanding of magnetic
dissipation.

First of all, the basic form of the scaling law
(\ref{nordlund-eq-scaling}) follows from first principles, and
agrees with previously proposed scaling laws (\cite{Parker83d},
\cite{vanBall86}).  In addition, it adds a prediction for the
crucial inclination factor, for which Parker (1983) and van
Ballegooijen (1986) had to make arbitrary assumptions.

Secondly, the \inx{scaling law}  is robust in that the crucial
inclination factor is bracketed from above and below.  It is
bracketed from below because in general current sheets do not
develop until the \inx{local winding number} is of the order of unity.
It is bracketed from above, because a local winding number much
larger than unity inevitably leads to instabilities that rapidly
dissipate the surplus magnetic energy.

Because of the spontaneous formation of a hierarchy of current
sheets the scaling law is also robust against changes of the
magnetic Reynolds number.  If anything, an increased Reynolds number
could in principle lead to an {\em increase} of the magnetic
dissipation because, as pointed out by Parker (1988), if one
assumes that a reduction of the magnetic diffusivity initially
leads to a reduction of the magnetic dissipation the consequence
is only that the boundary work for a while exceeds the dissipation,
which leads to an increase of the average inclination and hence
to a further increase of the boundary work.  When the magnetic
dissipation eventually comes into balance with
the boundary work again it happens at a {\em higher} level than
before.

But in practice, an increase of the magnetic Reynolds number
$\Rm$ just leads to an extension of the hierarchy of
electrical current sheets to smaller scales, which makes it
possible to dissipate at the same rate even without increasing
the average angle of inclination at the boundaries.  Any claim to
the contrary must be accompanied by a demonstration that it is
possible to sustain a distribution of winding number that is
substantially wider than unity at high $\Rm$.

The scaling law (\ref{nordlund-eq-scaling}) is furthermore
consistent with \inx{scaling laws derived from observations}, in that
it predicts magnetic dissipation to scale as $L^{-2}$, in
agreement with Porter \& Klimchuk (1995).  Also, since the solar
photosphere has a very intermittent distribution of magnetic
field strength $B$, where $B$ is either very weak or of the
order of 1 kG, the average heating is predicted to scale roughly as the
\inx{magnetic surface filling factor}.  This again agrees with the
\inx{observed scaling of coronal heating} (\cite{Fisher+98}).

\begin{acknowledgments}
This work was supported in part by the Danish National Research Foundation
through its establishment of the Theoretical Astrophysics Center.
\end{acknowledgments}

%% Just before the bibliography please list index entries for all
%% authors that appear in your bibliography:

%% \anxx{Kawahara, G.} \anxx{Kida, S.} \anxx{Yanase. S} \anxx{Tanaka, M.}
%% \anxx{Dopazo, C.}

\anxx{Aschwanden, M. J.} \anxx{Schrijver, C. J.}
\anxx{Fisher, G. H.} \anxx{Longcope, D. W.} \anxx{Metcalf, T. R.} \anxx{Pevtsov, A. A.}
\anxx{Galsgaard, K.} \anxx{Nordlund, {\AA}.}
\anxx{Gomez, D. O.} \anxx{Dmitruk, P. A.} \anxx{Milano, L. J.}
\anxx{Gudiksen, B.}
\anxx{Heyvaerts, J.} \anxx{Priest, E. R.}
\anxx{Longcope, D. W.} \anxx{Sudan, R.N.}
\anxx{Miki{\'c}, Z.} \anxx{Schnack, D. D.} \anxx{van Hoven, G.}
\anxx{Stein, R. F.} \anxx{Ludwig, H.-G.} \anxx{Rieutord, M.}
\anxx{Parker, E. N.}
\anxx{Porter, L. J.} \anxx{Klimchuk, J. A.}
\anxx{Sturrock, P. A.} \anxx{Uchida, Y.}
\anxx{van Ballegooijen, A. A.}

\end{document}